\documentstyle[12pt]{article}
\begin{document}
\title{{\bf PROBABILITIES DON'T MATTER}
\thanks{Alberta-Thy-28-94, gr-qc/9411004, to be published in {\em
Proceedings of the 7th Marcel Grossmann Meeting on General
Relativity}, eds. M. Keiser and R.T. Jantzen (World Scientific,
Singapore 1995).}}
\author{
Don N. Page
\thanks{Internet address:
don@phys.ualberta.ca}
\\
CIAR Cosmology Program, Institute for Theoretical Physics\\
Department of Physics, University of Alberta\\
Edmonton, Alberta, Canada T6G 2J1
}
\date{(1994 Nov. 25)}

\maketitle
\large
\begin{abstract}
\baselineskip 14.7 pt

	It is suggested that probabilities need not apply at all to
matter in the physical world, which may be entirely described by the
amplitudes given by the quantum mechanical state.  Instead,
probabilities may apply only to conscious perceptions in the mental
world.  Such perceptions may not form unique sequences that one could
call individual minds.
\\
\\
\end{abstract}
\normalsize
\baselineskip 14.7 pt

\section{Probabilities in Quantum Mechanics}

\hspace{.25in}One of the most mysterious aspects of quantum mechanics
is its usual probabilistic interpretation.  There is first the
uncertainty of which amplitudes should be squared to get
probabilities.  Then there is the even deeper puzzle of what the
resulting probabilities mean.

	For example, one viewpoint on the first question is that
whenever a measurement is made, the amplitude for each
macroscopically-distinct outcome should be squared to get a
probability.  (More precisely, one takes a complete set of orthogonal
projection operators, each representing one of the
macroscopically-distinct outcomes.  One projects the wavefunction by
one of these projection operators to get a reduced wavefunction.
Then one takes the inner product of this reduced wavefunction with
itself---i.e., ``squares the amplitude''---to get the probability of
the corresponding outcome.  This probability is the same as the
expectation value of the projection operator in the quantum state
given by the original wavefunction.)

	A broader viewpoint is that the expectation value of any
projection operator is a probability for the corresponding ``event.''
An even broader viewpoint is that one can square the amplitude given
by projecting the wavefunction not just by one, but by a whole
sequence of (possibly noncommuting) projection operators representing
a ``history'' or sequence of ``events.''  (For the resulting
probabilities to obey the sum rules under a coarse-graining of the
projection operators, the sequences must obey certain consistency
conditions \cite{Griffiths,Omnes}.)  One can extend this viewpoint,
of assigning probabilities to ``consistent histories,'' yet further
to the viewpoint that one can project the wavefunction by sums of
sequences of projection operators that represent coarser-grained
histories.  (Then one needs ``decoherence'' conditions for the
resulting probabilities to obey the sum rules for ``decohering
histories'' \cite{GMH,H,GMH2}.)  An even further extension is the
viewpoint that probabilities are the real parts of the expectation
values of sums of sequences of projection operators, whenever these
obey a ``linear positivity'' condition of being nonnegative, giving
probabilities for ``linearly positive histories'' that automatically
obey the sum rules \cite{GP}.

	Besides this sample of the wide variety of viewpoints of what
quantities should be assigned probabilities in quantum mechanics,
there is the enigma of how to interpret the resulting probabilities.
One attitude is that a unique one of the possible ``events'' or
``histories'' actually occurs, with the probability assigned by
quantum mechanics, and that the other possibilities do not.  This
attitude still leaves open the question of whether the
``probability'' is a ``frequency'' in an ensemble of actual worlds,
or whether it is a ``propensity'' in a single world and what that
could mean.  If the latter, what is it that actually makes the choice
of the actual ``event'' or ``history'' from those potentially
possible?

	Another attitude is that all of the possible ``events'' or
``histories'' with nonzero probabilities actually occur, but with
measures proportional to the probabilities.  This ``many-worlds''
interpretation \cite{E} is very similar to the frequency
interpretation in a ensemble of actual worlds mentioned above, but it
need not have the implication that our present world is a single
member of the ensemble that has a definite (albeit unpredictable)
future.

	In any of these approaches in which one or all of the
possibilities are actualized, there is the further question of which
set of possibilities is singled out.  In general, there are many
different allowed sets of possibilities (e.g., different sets of
orthogonal projection operators, or of sequences of projection
operators, that each add up to the identity operator).  If only one
``event'' or only one ``history'' actually occurs, there must be
mysterious choices both of the set of possibilities and of the single
actual element within that chosen set.  If, on the other hand, all of
the possible ``events'' or ``histories'' with nonzero probabilities
in a given set of possibilities actually occur, there must apparently
still be a mysterious choice of this set of possibilities out of the
family of all such sets of possibilities.

	One conceivable answer is that all possibilities in all sets
of possibilities actually occur, with measures given by the
probabilities deduced from quantum mechanics in one of the ways
discussed above.  Since, for a normalized quantum state, these
probabilities are designed to add up to unity for a single set of
possibilities, the measures will sum to more than unity when one adds
up all the sets.  In fact, typically the number of allowed sets of
possibilities is infinite (e.g., even for a two-state spin-half
system, there is a rank-one projection operator for each direction in
space and hence an infinite number of such projection operators).
This means that the sum of the measures for all possibilities
typically is unnormalizable, which may lead to problems.  (These
problems may be avoidable.  For example, in the spin-half system, the
sum of the quantum probabilities for all of the infinite number of
possibilities for the spin direction is infinite, but instead of
simply adding these quantum probabilities, one can divide by the
total solid angle of the unit sphere of spin directions to get a
probability density which can then be integrated over a nonzero solid
angle to get a normalized second-order probability that the spin is
in a direction within that solid angle.)

\section{Sensible Quantum Mechanics}

\hspace{.25in}Here I shall propose instead that probabilities not be
applied at all to the physical world (the ``matter'' of the title),
which is instead to be described entirely by the quantum amplitudes
of its wavefunction (or by the elements of its density matrix, or
possibly by a more general description, such as a $C^*$-algebra
state).  I suggest instead that probabilities apply only to the
mental world of conscious perceptions.  In this viewpoint, and in a
certain loose manner of speaking in which ``mind'' is this whole
mental world rather than a sequence of perceptions, probabilities are
only in the mind.

	Consider the set of all possible perceptions $p$, which I
shall call the mental world $M$.  By a perception, I mean all that
one is consciously aware or consciously experiencing at one moment.
This is what Lockwood, in a book expressing what seems to be ideas
mostly concordant with mine \cite{Lo}, calls a ``phenomenal
perspective'' or ``maximal experience.''  In another way of putting
it, my $p$ denotes a total ``raw feel'' that one has at once.  It can
include components such as a visual sensation, an auditory sensation,
a pain, a conscious memory, a conscious impression of a thought or
belief, etc., but it does not include a sequence of more than one
immediate perception that in other proposals might be considered to
be strung together to form a stream of consciousness of an individual
mind.

	Suppose that there is a basic measure on the mental world
which weights each perception equally, which I shall denote by $dp$
without implying a choice between the conceptual possibilities that
the space of perceptions is discrete (in which case $dp$ simply
counts the number of perceptions) or continuous (in which case $dp$
might represent some basic multi-dimensional integration measure or
volume element).  Now I shall postulate that our actual world does
not have all possible perceptions occurring equally, but instead that
there is a nontrivial measure $m(p)dp$, the nonnegative real scalar
function or measure density $m(p)$ times the basic measure, for the
perceptions in our actual world.  Thus I get the following basic
assumption:

	{\bf Measure Axiom for Perceptions}:  There is a nontrivial
measure on the space of (maximal) perceptions $p$, namely $m(p)$
times the basic measure $dp$ that weights each perception equally.

	Because of the fact that our perceptions seem to be more
simply explained by assuming that they are related to a physical
world, I shall assume a principle of psycho-physical parallelism,
that the nontrivial part $m(p)$ of the measure for perceptions is a
functional of the properties of the physical world.

	For example, if the physical world were represented by a
single classical trajectory in some phase space, it might be natural
to assume that $m(p)$ has a form of a sum of Kronecker deltas or an
integral of a discrete set of Dirac delta function distributions over
sequences of perceptions that a set of conscious beings has as the
point moves along the trajectory in phase space.  In other words,
each point in phase space might naturally be assigned a discrete set
of conscious perceptions, one for each conscious being whose point in
its phase space is derived from the point in the phase space of the
entire system (e.g., the universe).  Even in this classical model,
there can be many perceptions at one time, but each is ascribed to a
different conscious being.

	If the physical world is represented by a quantum state that
has no preferred classical trajectories (such as do occur in Bohm's
version of quantum mechanics \cite{B}, but which I shall not further
consider here), then it seems unnatural to assume that $m(p)$ is
completely concentrated on a discrete sequence of trajectories, one
for each conscious observer.  Instead, in view of the linearity of
quantum mechanics, I propose the following basic assumption for
interpreting quantum mechanics `sensibly,' i.e., in terms of
sensations or perceptions:

	{\bf Sensible Quantum Axiom}: Each $m(p)$ is given by the
expectation value of a corresponding positive perception or ``maximal
experience'' operator $E(p)$ in the quantum state of the universe.
As a formula,
	\begin{equation}
	m(p) = \langle E(p) \rangle = \langle\psi|E(p)|\psi\rangle =
Tr(E(p)\rho),
	\label{eq:1}
	\end{equation}
where the third expression applies if the quantum state is
represented by the wavefunction or pure state $|\psi\rangle$, and the
fourth expression applies if the quantum state is represented by the
statistical operator or density matrix $\rho$.  (The second
expression can apply in more general situations, such as in
$C^*$-algebra.)

	In this framework, which I shall call Sensible Quantum
Mechanics, the quantum state of the universe is fixed, in the
Heisenberg picture I am using, and never collapses or changes to
another state, so the von Neumann intervention {\bf 1} \cite{vN} is
assumed never to occur.  Neither is there assumed to be any
nonlinearity in quantum mechanics when consciousness is involved, as
Wigner proposed \cite{W}.  If the quantum state and all the
perception operators $E(p)$ are known, one can in principle calculate
from Eq.~(\ref{eq:1}) the measure density $m(p)$ for all perceptions.
(Of course, I am not competent to give these essential elements, so
the present proposal is a framework, on the level of other frameworks
or interpretations of quantum mechanics, rather than a complete
theory.)  Since all maximal perceptions $p$ with $m(p) > 0$ really
occur in this framework, it is completely deterministic if the
quantum state and the $E(p)$ are determined:  there are no random or
truly probabilistic elements in this framework of Sensible Quantum
Mechanics.

	Thus Sensible Quantum Mechanics has no need for any axiom of
what is typically called ``measurement'' \cite{vN}, or what Unruh
calls ``determination'' \cite{U} to distinguish this hypothetical
process from the physical measurement interactions that are encoded
in the quantum state and the structure of the operators.  In
particular, there are no probabilistic results of such
``determinations.''

	Nevertheless, because the framework has measures for
perceptions, one can readily use them to calculate quantities that
can be interpreted as conditional probabilities.  One can consider
sets of perceptions $A$, $B$, etc., defined in terms of properties of
the perceptions.  For example, $A$ might be the set of perceptions in
which there is a feeling that the universe is approximately described
by a Friedman-Robertson-Walker model, and $B$ might be the set of
perceptions in which there is a feeling that the universe is
approximately described by a Friedman-Robertson-Walker model with an
age (at the perceived time) between ten and twenty billion years.  By
summing or integrating $m(p)dp$ over the sets $A,\;B,\;A\cap B\; (=B$
in the example here), etc., one can get corresponding measures
$m(A),\;m(B),\;m(A\cap B)$, etc.  Then one can interpret
	\begin{equation}
	P(B|A)\equiv m(A\cap B)/m(A)
	\label{eq:2}
	\end{equation}
as the conditional probability that the perception is in the set $B$,
given that it is in the set $A$.  In our example, this would be the
conditional probability that a perception including the feeling that
the universe is approximately described by a
Friedman-Robertson-Walker model, also has the feeling that at the
time of the perception the age is between ten and twenty billion
years.

	An analogue of this conditional ``probability'' is the
conditional probability that a person in 1994 is the Queen of
England.  If we consider a model of all the five to six billion
people, including the Queen, that we agree to consider as our
contemporary humans on Earth in 1994, then at the basic level of this
model the Queen certainly exists in it; there is nothing random or
probabilistic about her existence.  But if the model weights each of
the five to six billion people equally, then one can in a manner of
speaking say that the conditional probability that one of these
persons is the Queen is somewhat less than $2\!\!\times\!\!
10^{-10}$.  I am proposing that it is in the same manner of speaking
that one can assign conditional probabilities to sets of perceptions,
even though there is nothing truly random about them at the basic
level.

	When one's perceptions include feelings of belief about what
is ascribed to be external events or histories (e.g., results of
experiments in the physical world rather than in the mind), and when
it is believed that these beliefs are an accurate representation of
some aspects of those ascribed-to-be-external events or histories, it
is tempting to the theorist to interpret the conditional
probabilities of the perceptions as giving conditional probabilities
for those aspects of the ascribed-to-be-external events or histories.
One may even go further and develop formalisms for directly
calculating probabilities of such events or histories.  With this
viewpoint one can say that the historical development of quantum
mechanics has been fruitful (or more strictly, I have a feeling that
what I perceive to have been its historical development was
fruitful), but it has left unexplained which events or histories are
to be assigned probabilities and what those probabilities mean.

	Thus I am proposing that at the basic level, probabilities
(or, more strictly, measures) have meaning only for perceptions in
the mental world and should not be assigned to anything (e.g., to
events or histories) in the physical world of matter.  In this sense
``probabilities don't matter.''

	On the other hand, I am not saying that it is forbidden to
assign conditional ``probabilities'' to events and/or histories in
the physical world.  They can be much simpler to calculate than those
in the mental world given by Eqs. (\ref{eq:1}) and (\ref{eq:2}),
since we don't know what the perception operators $E(p)$ are, and
these physical ``probabilities'' may often give good approximations
for the mental probabilities.  Indeed, I have been happy to help play
the game of broadening the scope of histories to which one can assign
mathematically-consistent probabilities \cite{GP}.  However, I am now
proposing that these probabilities are not fundamental and need not
be added to complicate the basic ontology of a measurable set of
perceptions (which I have been calling the mental world) and a
quantum state of the universe (which I have been calling the physical
world), with the measure and interpretation given entirely by the
Measure Axiom for Perceptions and the Sensible Quantum Axiom above.

	Goldstein has pointed out \cite{Gpri} that one can simplify
the ontology and avoid the assumption of a basic measure $dp$ on the
mental world by replacing the measure density $m(p)$ for single
perceptions $p$ with a measure $\mu(S)$ on sets $S$ of perceptions
and by replacing the perception operators $E(p)$ with a
positive-operator-valued measure, say $A(S)$.  This alternative
formulation is given in \cite{Winn,SQM}.  Another way to minimize the
number of independent entities is to postulate that the basic measure
$dp$ is the volume element of a Riemannian metric
	\begin{equation}
	g_{ij}dp^idp^j =
Tr\{[E(p^i+dp^i)-E(p^i)][E(p^j+dp^j)-E(p^j)]\},
	\label{eq:2b}
	\end{equation}
if this is finite and nondegenerate, so that the basic measure is
determined by the perception operators themselves.

\section{Perceptions rather than Minds}

\hspace{.25in}Another point I should emphasize is that in Sensible
Quantum Mechanics, the set of all perceptions is basic, but not any
higher power of this set.  In other words, perceptions and the
measure $m(p)dp$ on them are basic, but not $n$-tuples of
perceptions, or measures on $n$-tuples of perceptions.  Thus, for
example, there is no fundamental notion of a correlation between
individual complete perceptions given by any measure.  (On the other
hand, if a perception can be broken up into component parts, say $A$
and $B$, there can be a correlation between the parts, in the sense
that the measure $m(A\cap B)$ for all perceptions containing the part
$A$ and the part $B$ need not be the same as $m(A)m(B)$, the measure
for all perceptions containing $A$ times the measure for all
perceptions containing $B$.  The enormous structure in a single
perception seems to suggest that such correlations within perceptions
are highly nontrivial, but I see no evidence for a nontrivial
correlation between maximal perceptions, since no two different
maximal perceptions can be perceived together.)

	Furthermore, Sensible Quantum Mechanics postulates no
fundamental equivalence relation on the set of perceptions.  For
example, the measure gives no way of classifying different
perceptions as to whether they belong to the same conscious being
(e.g., at different times) or to different conscious beings.  The
only such classification would be by the content (including the {\it
qualia}) of the perceptions themselves, which distinguish the
perceptions, so that no two different perceptions, $p\neq p'$, have
the same content.  Based upon my own present perception, I find it
natural to suppose that perceptions that could be put into the
classification of being alert human perceptions have such enormous
structure that they could easily distinguish between all of the
$10^{11}$ or so persons that are typically assigned to our history of
the human race.  However, I doubt that in a fundamental sense there
is any absolute classification that uniquely distinguishes each
person in all circumstances.  Therefore, in the present framework
perceptions are fundamental, but persons (or individual minds) are
not, although they certainly do seem to be very good approximate
entities that I do not wish to deny.  The concept of persons and
minds certainly occurs in some sense as part of the {\it content} of
my present perception, even if there is no absolute definition of it
in the framework of Sensible Quantum Mechanics itself.

	In this way the framework of Sensible Quantum Mechanics
proposed here is a particular manifestation of Hume's ideas
\cite{Hume}, that ``what we call a {\it mind}, is nothing but a heap
or collection of different perceptions, united together by certain
relations, and suppos'd, tho' falsely, to be endow'd with a perfect
simplicity and identity'' (p. 207), and that the self is ``nothing
but a bundle or collection of different perceptions'' (p. 252).  As
he explains in the Appendix (p. 634), ``When I turn my reflexion on
{\it myself}, I never can perceive this {\it self} without some one
or more perceptions; nor can I ever perceive any thing but the
perceptions.  'Tis the composition of these, therefore, which forms
the self.''  (Here I should note that what Hume calls a perception
may be only one component of the ``phenomenal perspective'' or
``maximal experience'' \cite{Lo} that I have been calling a
perception $p$, so one $p$ can include ``one or more perceptions'' in
Hume's sense.)

	Furthermore, each perception operator need not have any
precise location in either space or time associated with it, so there
need be no fundamental place or time connected with each perception.
Indeed, Sensible Quantum Mechanics can easily survive a replacement
of spacetime with some other structure (e.g., superstrings) as more
basic in the physical world.  Of course, the contents of a perception
can include a sense or impression of the time of the perception, just
as my present perception at the perceived time of writing this
includes a feeling that it is now 1994 A.D., so the set of
perceptions $p$ must include perceptions with such beliefs, but there
need not be any precise time in the physical world associated with a
perception.  That is, perceptions are `outside' physical spacetime
(even if spacetime is a fundamental element of the physical world,
which I doubt).

	As a consequence of these considerations, there are no unique
time-sequences of perceptions to form an individual mind or self in
Sensible Quantum Mechanics.  In this way the present framework
appears to differ from those proposed by Squires \cite{Sq}, Albert
and Loewer \cite{A}, and Stapp \cite{St}.  (Stapp's also differs in
having the wavefunction collapse at each ``Heisenberg actual event,''
whereas the other two agree with mine in having a fixed quantum
state, in the Heisenberg picture, which never collapses.)  Lockwood's
proposal \cite{Lo} seems to be more similar to mine, though he also
proposes (p. 232) ``a continuous infinity of parallel such streams''
of consciousness, ``{\it differentiating} over time,'' whereas
Sensible Quantum Mechanics has no such stream as fundamental.  On the
other hand, later Lockwood \cite{Lo2} does explicitly repudiate the
Albert-Loewer many-minds interpretation, so there seems to me to be
little disagreement between Lockwood's view and Sensible Quantum
Mechanics except for the detailed formalism and manner of
presentation.  Thus one might label Sensible Quantum Mechanics as the
Hume-Everett-Lockwood-Page (HELP) interpretation, though I do not
wish to imply that these other three scholars, on whose work my
proposal is heavily based, would necessarily agree with my present
formulation.

	Of course, the perceptions themselves can include components
that  seem to be memories of past perceptions or events.  In this way
it can be a very good approximation to give an approximate order for
perceptions whose content include memories that are correlated with
the contents of other perceptions.  It might indeed be that the
measure (or measure density) $m(p)$ for perceptions including
detailed memories is rather heavily peaked around approximate
sequences constructed in this way.  But I would doubt that either the
content of the perceptions or the measure on the set of perceptions
would give unique sequences of perceptions that one could rigorously
identify with individual minds.

	Because the physical state of our universe seems to obey the
second law of thermodynamics, with growing correlations in some
sense, I suspect that the measure density $m(p)$ may have rather a
smeared peak (or better, ridge) along approximately tree-like
structures of branching sequences of perceptions, with perceptions
further out along the branches having contents that includes memories
that are correlated with the present-sensation components of
perceptions further back toward the trunks of the trees.  This is
different from what one might expect from a classical model with a
discrete number of conscious beings, each of which might be expected
to have a unique sharp sequence or non-branching trajectory of
perceptions.  In the quantum case, I would expect that what are
crudely viewed as quantum choices would cause smeared-out
trajectories to branch into larger numbers of smeared-out
trajectories with the progression of what we call time.  If each
smeared-out trajectory is viewed as a different individual mind, we
do get roughly a ``many-minds'' picture that is analogous to the
``many-worlds'' interpretation \cite{E}, but in my framework of
Sensible Quantum Mechanics, the ``many minds'' are only approximate
and are not fundamental as they are in the proposal of Albert and
Loewer \cite{A}.  Instead, Sensible Quantum Mechanics is a
``many-perceptions'' or ``many-sensations'' interpretation.  One
might label it philosophically as Mindless Sensationalism.

	Even in a classical model, if there is one perception for
each conscious being at each moment of time in which the being is
conscious, the fact that there may be many conscious beings, and many
conscious moments, can be said to lead to a ``many-perceptions''
interpretation.  However, in Sensible Quantum Mechanics, there may be
vastly more perceptions, since they are not limited to a discrete set
of one-parameter sharp sequences of perceptions, but occur for all
perceptions $p$ for which $m(p)$ is positive.  In this way a quantum
model may be said to be even ``more sensible'' than a classical
model.

	One might fear that the present attack on the assumption of
any definite notion of a precise identity for persons or minds as
sequences of perceptions would threaten human dignity.  Although I
would not deny that I feel that it might, I can point out that on the
other hand, the acceptance of the viewpoint of Sensible Quantum
Mechanics might increase one's sense of identity with all other
humans and other conscious beings.  Furthermore, it might tend to
undercut the motivations toward selfishness that I perceive in myself
if I could realize in a deeply psychological way that what I normally
anticipate as my own future perceptions are in no fundamental way
picked out from the set of all perceptions.  (Of course, what I
normally think of as my own future perceptions are presumably those
that contain memory components that are correlated with the content
of my present perception, but I do not see logically why I should be
any more concerned about trying to make such perceptions happy than
about trying to make perceptions happy that do not have such
memories:  better to do unto others as I would wish they would do
unto me.)  Lockwood \cite{Lpri} informs me that Parfit \cite{Par} has
drawn similar conclusions from a Humean view.

\section{Properties of Perception Operators}

\hspace{.25in}Although no one is competent to give the complete set
of perception operators $E(p)$, one can speculate about some of their
properties.  In this speculation, a theoretical physicist such as
myself would like to be guided by the principles of simplicity and of
agreement with observations.  Both are difficult, the former because
we do not know all that is logically possible and have a measure of
the simplicity of the different possibilities, and the latter because
we do not have direct access to more than one perception at once.

	On the former principle, it is because of simplicity that I
do not stop at the Measure Axiom for Perceptions but also postulate a
physical quantum state and a set of perception operators $E(p)$ which
give the measure density $m(p)$ by the Sensible Quantum Axiom.  If
one has the correct $m(p)$ (as well as the basic measure $dp$, which
is a separate element from the set of perceptions if they are not
discrete), the Measure Axiom for Perceptions is logically sufficient
for describing a measured set of perceptions.  It might seem to be
complicating the theory to add a physical quantum state and a set of
perception operators $E(p)$, but I believe that this structure of a
postulated physical world can give a simpler explanation of $m(p)$
than just giving it directly without this explanation in terms of a
postulated physical world.  In this way the whole of the mental world
and the physical world can be simpler than just the mental world
considered by itself.  (One might also raise the reverse question of
whether the whole is simpler than the physical world alone, by which
I mean an alternative logically possible world in which all $E(p)$
are zero, so that in it the quantum state is the same as in ours, but
no conscious perceptions occur.)

	On the latter principle, the only agreement with observations
that one can impose is the assumption that one's perception be not
too atypical, i.e., that it have not too low a measure density
$m(p)$.  For example, if both the basic measure $dp$ and the quantum
measure $m(p)dp$ were integrable, over the set of all perceptions
$p$, to finite numbers, say $b$ and $q$ respectively, then one can
ask that one's particular perception not have $m(p)\ll q/b$, the
latter being the average of $m(p)$ over all perceptions.
Unfortunately, I see no reason why a simple theory should make either
of these integrals finite, since almost any finite number is more
complex than infinity.  (Perhaps the fact that my present perception
seems to have a large but not infinite amount of information in it is
evidence that the simplest complete theory is not extremely simple,
since I would expect an extremely simple theory to make typical
perceptions have either an extremely small or an infinite amount of
information.)

	Perhaps a more realistic approach one can make toward
agreement with observations is to assume that the measure density
$m(p)$ for one's perception is not much lower than the measure
density for slightly different perceptions.  For example, if one has
a perception $p$ of having made a certain quantum measurement $n$
times and having gotten $m$ positive results, one can imagine another
perception $p'$ which is similar except for perceiving $m'$ positive
results.  Then one would like $m(p)$ to be not much lower than
$m(p')$.  If the measures (or measure densities) $m(p)$ and $m(p')$
of the perceptions $p$ and $p'$ can to a good approximation be
replaced by the quantum-derived measures $\tilde{m}(r)$ and
$\tilde{m}(r')$ for the respective perceived results $r$ and $r'$ in
the physical world, one can check whether $\tilde{m}(r)$ is not much
lower than $\tilde{m}(r')$.  In such cases there is considerable
experimental evidence that the ordinary quantum predictions are
consistent with observations if $\tilde{m}(r)$ is the expectation
value of a projection operator $P(r)$ for the physical result $r$.

	Similarly, it might be natural to assume that each perception
measure density $m(p)$ is given by the expectation value of a
projection operator, say one that projects onto the brain states that
cause the perception, if indeed the perception is caused by various
brain states.  However, before making this specific assumption, let
me make some weaker postulates that one could add to Sensible Quantum
Mechanics to make it more restrictive and yet have a more specific
content:

	{\bf Hypothesis A}:  The expectation value of each $E(p)$ has
a constant maximum value, say unity, in the set of all normalized
quantum states.

	Assuming that the quantum state is normalizable (perhaps an
overly restrictive assumption), Hypothesis A and the resulting
Sensible Quantum Mechanics A would have $0 \leq m(p) \leq 1$ for all
$p$, and there would exist a normalizable quantum state for each $p$
such that the corresponding $m(p)$ would be unity in that state.

	Without some such restriction like that, one could leave all
the explanation of $m(p)$ in the operators $E(p)$ rather than in the
quantum state.  For example, one could get $m(p)$ to be any positive
function whatsoever simply by choosing $E(p)$ to be that function
times the identity operator, which would make $m(p)$ independent of
the state (so long as the state is normalized so that the expectation
of the identity operator is unity; other normalizations would change
the scale of the measure but would leave the conditional
probabilities of Eq.~(\ref{eq:2}) unchanged).  We would like to
assume that instead the $E(p)$ are restricted so the explanation for
the $m(p)$ lies largely in the quantum state.

	Now Hypothesis A is still not highly restrictive, so one may
wish to look for more restrictions on the operators $E(p)$.  For
example, one may be motivated by the consistent or decohering
histories approaches \cite{Griffiths,Omnes,GMH,H,GMH2} to assume that
perceptions are connected with histories and so perhaps make the
following assumption:

	{\bf Hypothesis B}:  Each $E(p)$ has the form
$E(p)=C^{\dagger}C/max(C^{\dagger}C)$, where
$C=P^{(n)}P^{(n-1)}\cdots P^{(2)}P^{(1)}$ with the integer $n$ and
the projection operators $P^{(i)}$ all depending on the perception
$p$, and where $max(C^{\dagger}C)$ is the maximum expectation value
of $C^{\dagger}C$ in any normalized quantum state.

	The denominator in the expression for $E(p)$ in Hypothesis B
is chosen so that Hypothesis B is a special case of Hypothesis A, but
one could also consider an alternative Hypothesis B' in which the
denominator is omitted.  One could also consider generalizing
Hypothesis B or B' to B* or B*' respectively, in which $C$ is a sum
of sequences of projection operators, as is allowed in decohering
histories \cite{GMH}.

	It is certainly logically possible that perceptions might
depend on histories rather than events that one could consider
localized on hypersurfaces of constant time if the physical world has
such a time.  However, as a previous advocate of the `marvelous
moment' approach to quantum mechanics in which only quantities on one
such hypersurface can be tested \cite{P}, I find it more believable
to assume that perceptions are caused by brain states which could be
at one moment of time if there are such things in the physical world.
The generalization of this hypothesis to the case in which there may
not be a well-defined physical time leads me to make the following
restriction of Hypothesis B or B' to the case in which the integer
$n$ is always 1:

	{\bf Hypothesis C}:  $E(p)=P(p)$, a projection operator that
depends on the perception $p$.

	Hypothesis C appears to be a specific mathematical
realization of part of Lockwood's proposal \cite{Lo} (p. 215), that
``a phenomenal perspective [what I have here been calling simply a
perception $p$] may be equated with a shared eigenstate of some
preferred (by consciousness) set of compatible brain observables.''
Here I have expressed the ``equating'' by Eq.~(\ref{eq:1}), and
presumably the ``shared eigenstate'' can be expressed by a
corresponding projection operator $P(p)$.

	I should also emphasize that if the same conscious perception
is produced by several different orthogonal ``eigenstates of
consciousness'' (e.g., different states of a brain and surroundings
that give rise to the same perception $p$), then in Hypothesis C the
projection operator $P(p)$ would be a sum of the corresponding
rank-one projection operators and so would be a projection operator
of rank higher than unity (perhaps even infinite), which is what I
would expect.  On the other hand, if $E(p)$ were a sum of
noncommuting projection operators corresponding to nonorthogonal
states, or a weighted sum of projection operators with weights
different from unity, then generically $E(p)$ would not have the form
of a projection operator $P(p)$.

	If one has a constrained system, such as a closed universe in
general relativity, the quantum state may obey certain constraint
equations, such as the Wheeler-DeWitt equations.  The projection
operators $P(p)$ of perception in Hypothesis C may not commute with
these constraints, in which case they may give technically
`unphysical' states when applied to the quantum state.  But so long
as their expectation values can be calculated, that is sufficient for
giving the perception measure density $m(p)$.  What it means is that
in Hypothesis C, the perception operators should be considered as
projection operators in the space of unconstrained states, even
though the actual physical state does obey the constraints.

	Alternatively, if one wishes to write the perception
operators $E(p)$ as operators within the space of constrained states,
Hypothesis C could be modified to the following assumption to give
perception operators $E(p)$ that commute with the constraints and so
keep the state `physical':

	{\bf Hypothesis \~{C}}:  $E(p)=P_CP(p)P_C$, where $P_C$ is
the projection operator within the space of unconstrained states that
takes any state to the corresponding constrained state, and $P(p)$ is
a projection operator in the space of unconstrained states that
depends on the perception $p$.

	One could obviously alternatively insert the projection
operator $P_C$ before and after the perception operators of
Hypothesis B, B', B*, or B*' to get Hypothesis \~{B}, \~{B}', \~{B}*,
or \~{B}*' respectively.

	One can also get something like Hypothesis \~{C}, say
Hypothesis \^{C}, even for unconstrained systems if they have
symmetries (e.g., the Poincar\'{e} symmetries of quantum field theory
in a classical Minkowski spacetime, though one would not expect these
symmetries to survive when one includes gravity), since one might
then expect that $E(p)$ should be invariant under the symmetry group
with elements $g$.  Then if one starts with a projection operator
$P(p)$ that is not invariant under the action of each group element,
say $P(p)\neq gP(p)g^{-1}$, then one might expect $E(p)$ to be
proportional to the sum or integral of $gP(p)g^{-1}$ over the group
elements $g$.  Unless all these different $gP(p)g^{-1}$'s are
orthogonal (which does not appear possible for a continuous symmetry
group), the resulting $E(p)$ will generically not be a projection
operator, but it can be said to have arisen from one, which is what I
would propose as the translation of the marvelous moment assumption
to Sensible Quantum Mechanics.

	If one can learn what the $E(p)$'s are, one can compare them
with the forms given by these hypotheses and thereby distinguish
between the consistent or decohering histories approaches and the
marvelous moment approach as I here propose they be applied to
conscious perceptions (if indeed any of them do).  Of course, either
of these approaches could be applied without inconsistency to
mathematical probabilities that one might wish to define in the
physical world, but in the present framework of Sensible Quantum
Mechanics, such probabilities are an unnecessary addition to the
ontology.

	I should emphasize that in no case am I assuming that the
$E(p)$'s commute for different perceptions, or that the sum or
integral of the $E(p)$'s over all perceptions is the identity
operator.  Neither am I assuming that the resulting expectation
values $m(p)$ in the particular quantum state of the universe are
normalized so that their sum or integral over all perceptions gives a
finite number $q$, or that this number is unity, though in any case
the conditional probabilities given by Eq.~(\ref{eq:2}) are
automatically normalized to give unity when summed over a complete
set of disjoint sets $B$ of perceptions.  Of course, if $q$ is
finite, one can simply rescale $E(p)$ to $e(p)=E(p)/q$, which
rescales $m(p)$ to $m(p)=m(p)/q$ that is normalized to give unity
when summed or integrated over all perceptions.  This rescaling
obviously leaves the conditional probabilities of Eq.~(\ref{eq:2})
invariant when $m(A\cap B)$ and $m(A)$ there are replaced by $m(A\cap
B)$ and $m(A)$ respectively.  On the other hand, I am sceptical that
the simplest consistent description of our universe will give a
normalizable $m(p)$ (finite $q$).

	If a perception operator $E(p)$ is a projection operator, and
the quantum state of the universe is represented by the pure state
$|\psi\rangle$, one can ascribe to the perception $p$ the pure
Everett ``relative state''
	\begin{equation}
	|p\rangle=\frac{E(p)|\psi\rangle}{\parallel
E(p)|\psi\rangle\parallel}
	=\frac{E(p)|\psi\rangle}{\langle\psi|E(p)E(p)|\psi\rangle^{1/2
}}.
	\label{eq:3}
	\end{equation}
Alternatively, if the quantum state of the universe is represented by
the density matrix $\rho$, one can associate the perception with a
relative density matrix
	\begin{equation}
	\rho_p=\frac{E(p)\rho E(p)}{Tr[E(p)\rho E(p)]}.
	\label{eq:4}
	\end{equation}
Either of these formulas can be applied when the perception operator
is not a projection operator, but then the meaning is not necessarily
so clear.

	If one has two perceptions $p$ and $p'$, one can calculate an
overlap fraction between them as
	\begin{equation}
	f(p,p')=\frac{\langle E(p)E(p')\rangle\langle
E(p')E(p)\rangle}
	{\langle E(p)E(p)\rangle\langle E(p')E(p')\rangle}.
	\label{eq:5}
	\end{equation}
If the quantum state of the universe is pure, this is the same as the
overlap probability between the two Everett relative states
corresponding to the perceptions:  $f(p,p')=|\langle p|p'\rangle|^2$.
Thus one might in some sense say that if $f(p,p')$ is near unity, the
two perceptions are in nearly the same one of the Everett ``many
worlds,'' but if $f(p,p')$ is near zero, the two perceptions are in
nearly orthogonal different worlds.  However, this is just a manner
of speaking, since I do not wish to say that the quantum state of the
universe is really divided up into many different worlds.  Thus I do
not wish to propose that $f(p,p')$ be interpreted as a fundamental
element of Sensible Quantum Mechanics.  In any case, one can be
conscious only of a single perception at once, so there is no way in
principle that one can test any properties of joint perceptions such
as $f(p,p')$.

\section{Quantum Field Theory Model}

\hspace{.25in}Although Sensible Quantum Mechanics transcends quantum
theories in which space and time are fundamental, and although I
believe that such theories will need to be transcended to give a good
theory of our universe, it might help to get a better feel for the
spacetime properties of perceptions by considering the context of
quantum field theory in an unquantized curved globally-hyperbolic
background spacetime in which spacetime points are unambiguously
distinguished by the spacetime geometry (so that the Poincar\'{e}
symmetries are entirely broken and one need not worry about
integrating over $gP(p)g^{-1}$'s to satisfy superselection rules for
energy, momentum, and/or angular momentum \cite{PW}).  This
simplified model might in some sense be a good approximation for part
of the entire quantum state of the universe in a correct theory if
there is one that does fit into the framework of Sensible Quantum
Mechanics and does give a suitable classical spacetime approximation.

	In the Heisenberg picture used in this paper, the quantum
state is independent of time (i.e., of a choice of Cauchy
hypersurface in the spacetime), but the Heisenberg equations of
evolution for the fundamental fields and their conjugate momenta can
be used to express the operators $E(p)$ in terms of the fields and
momenta on any Cauchy hypersurface.  The arbitrariness of the
hypersurface means that even in this quantum field theory with a
well-defined classical spacetime, and even with a definite foliation
of the spacetime by a one-parameter (time) sequence of Cauchy
hypersurfaces, there is no unique physical time that one can assign
to any of the perceptions $p$; they are `outside' time as well as
space.

	Furthermore, the operators $E(p)$ in this simplified model
are all likely to be highly nonlocal in terms of local field
operators on any Cauchy hypersurface, since quantum field theories
that we presently know do not seem to have enough local operators to
describe the complexities of an individual perception, unless one
considers high spatial derivatives of the field and conjugate
momentum operators.  However, for a given one-parameter (time)
sequence of Cauchy hypersurfaces, one might rather arbitrarily choose
to define a preferred time for each perception $p$ as the time giving
the Cauchy hypersurface on which the corresponding $E(p)$, if
expressed in terms of fields and momenta on that hypersurface, has in
some sense the smallest spatial spread at that time.

	For example, to give a tediously explicit {\it ad hoc}
prescription, on a Cauchy hypersurface labeled by the time $t$ one
might choose a point $P$ and a ball that is the set of all points
within a certain geodesic radius $r$ of the point.  Then one can
define an operator $E'(p;t,P,r)$ that is obtained from $E(p)$ written
in terms of the fields and conjugate momenta at points on the
hypersurface by throwing away all contributions that have any fields
or conjugate momenta at points outside the ball of radius $r$ from
the point $P$.  Now define the overlap fraction
	\begin{equation}
	F(p;t,P,r)=\frac{\langle E(p)E'(p;t,P,r)\rangle
	\langle E'(p;t,P,r)E(p)\rangle}
	{\langle E(p)E(p)\rangle
	\langle E'(p;t,P,r)E'(p;t,P,r)\rangle}.
	\label{eq:6}
	\end{equation}
(If both  $E(p)$ and $E'(p;t,P,r)$ were projection operators, and the
actual quantum state were a pure state, then $F$ would be the overlap
probability between the states obtained by projecting the actual
quantum state by these projectors and normalizing.)  If $E(p)$ is
nonlocal, this fraction $F$ will be small if the radius $r$ is small
but will be nearly unity if the radius $r$ is large enough for the
ball to encompass almost all of the Cauchy hypersurface.  For each
perception $p$, time $t$, and point $P$, one can find the smallest
$r$ that gives $F=1/2$, say, and call that value of the radius
$r(p;t,P)$.  Then one can find the point $P=P(p;t)$ on the
hypersurface that gives the smallest $r(p;t,P)$ on that hypersurface
for the fixed perception $p$ and call the resulting radius $r(p;t)$.
Finally, define the preferred time $t_p$ as the time $t$ for which
$r(p;t)$ is the smallest, and label that smallest value of $r(p;t)$
for the fixed perception $p$ as $r_p$.

	If the perception operator $E(p)$ for some human conscious
perception is not unduly nonlocal in the simplified model under
present consideration, and if the quantum state of the fields in the
spacetime has macroscopic structures that at the time $t_p$ of the
perception are fairly well localized (e.g., with quantum
uncertainties less than a millimeter, say, which would certainly not
be a generic state, even among states which give a significant $m(p)$
for the perception in question), one might expect that at this time
the ball within radius $r_p$ of the point $P(p;t_p)$ on the
hypersurface labeled by $t_p$ would be inside a human brain.  It
would be interesting if one could learn where the point $P(p;t_p)$ is
in a human brain, and what the radius $r_p$ is, for various human
perceptions, and how the location and size of this ball depends on
the perception $p$.

\section{Questions and Speculations}

\hspace{.25in}One can use the framework of Sensible Quantum Mechanics
to ask questions and make speculations that would be difficult
without such a framework.  I shall here give some examples, without
intending to imply that Sensible Quantum Mechanics itself, even if
true, would guarantee that these questions and speculations make
sense, but it does seem to allow circumstances in which they might.

	First, in the model of quantum field theory on a classical
spacetime with no symmetries, and with a quantum state having
well-localized human brains on some Cauchy hypersurface labeled by
time $t$, one might ask whether it is possible to have two quite
different perceptions, say $p$ and $p'$, in nearly the same Everett
world in the sense of having the $f(p,p')$ of Eq.~(\ref{eq:5}) near
unity, and giving $E(p)$ and $E(p')$ both with the same preferred
time $t_p=t$ and both localized (by the rather {\it ad hoc}
prescription above) in balls in the same brain.  In other words, can
one brain have two different (maximal) perceptions in the same world
at the same time, each not aware of the other?  Unless we are
solipsists, we generally believe this is possible for two separate
brains, but would one brain be sufficient?  Furthermore, if it is
possible, can the two balls (corresponding to $p$ and $p'$
respectively) be overlapping spatially, or need they be separate
regions in the brain?

	Second, one might ask whether and how the sum (or integral)
of the measures (or measure densities) $m(p)$ associated with an
individual brain region at the time $t$ depends on the brain
characteristics.  One might speculate that it might be greater for
brains that are in some sense more intelligent, so that in a crude
sense brighter brains have more perceptions.  This could explain why
you do not perceive yourself to be an insect, for example, even
though there are far more insects than humans.  To speculate even
further, it might imply that your perception is not atypical even if
you perceive yourself to be more intelligent than the average person,
as I predict that most of my readers do.  Of course, my prediction is
based on an assumed selection effect of those who read physics
papers, on my own sinful pride that leads me to assume that
physicists are brighter than average, and on a
psychologically-projected assumption that most people think they are
more intelligent than most people.  But if you really had good
reasons for believing that you were brighter than average, even
before reading this paper, you may not really be justified in any
feeling of atypicality; it might simply be that your perceptions,
like most perceptions in the measure $m(p)dp$, are associated with
brighter-than-average people.

	I should emphasize that even if this wild conjecture, which
is not an inevitable consequence of Sensible Quantum Mechanics (but
which can be considered under this framework in a way that might be
difficult under other frameworks), could be shown to be true, I would
not propose that a more intelligent person be assigned more value or
be given more political rights or privileges.  It would just say
that, when weighted by the quantum-mechanical operators $E(p)$ for
perceptions, it is conceivable that more intelligent people would
have the bulk of the weight rather than being unusual.  But being in
this newly-defined majority (if indeed it is the bulk of the weight)
should not confer more individual political rights or privileges,
just as with the weighting of numbers the majority of people who are
economically poor are not given more individual political rights or
privileges than the minority of people who are economically rich.

	Third, one might conjecture that an appropriate measure on
perceptions might give a possible explanation of why most of us
perceive ourselves to be living on the same planet on which our
species developed.  This observation might seem surprising when one
considers that we may be technologically near the point at which we
could leave Earth and colonize large regions of the Galaxy,
presumably greatly increasing the number of humans beyond the roughly
$10^{11}$ that are believed to have lived on Earth.  If so, why don't
we have the perceptions of one of the vast numbers of humans that may
be born away from Earth?  One answer is that some sort of doom is
likely to prevent this vast colonization of the Galaxy from happening
\cite{C,Le,N,G}, though these arguments are not conclusive
\cite{KKP}.  Although I would not be surprised if such a doom were
likely, I would na\"{\i}vely expect it to be not so overwhelmingly
probable that the probability of vast colonization would be as small
as is the presumably very small ratio of the total number of humans
who could ever live on Earth to those who could live throughout the
Galaxy if the colonization occurs.  Then, even though the
colonization may be unlikely, it may still produce a higher measure
for conscious perceptions of humans living off Earth than on it.

	However, another possibility is that colonization of the
Galaxy is not too improbable, but that it is mostly done by
self-replicating computers or machines who do not tolerate many
humans going along.  If the number of these dominate humans as
``intelligent'' beings, one might still have the question of why we
perceive ourselves as being humans rather than as being one of the
vastly greater numbers of such machines.  But the explanation might
simply be that the {\it weight} of conscious perceptions (the sum or
integral of the $m(p)$'s corresponding to the type of perceptions
under consideration) is dominated by human perceptions, even if the
{\it number} of ``intelligent'' beings is not.  In other words, human
brains may be much more efficient in producing conscious perceptions
than the kinds of self-replicating computers or machines which may be
likely to dominate the colonization of the Galaxy.  If such machines
are more ``intelligent'' than humans in terms of
information-processing cababilities and yet are less efficient in
producing conscious perceptions, our perceptions of being human would
suggest that the measure of perceptions is not merely correlated with
intelligence.

	It might be tempting to take the observations that these
speculations might explain (your perception of yourself as human
rather than as insect or even possibly as more intelligent than
average humans, if my prediction of that is correct, and our
perception of ourselves as humans on our home planet) as evidence
tending to support the speculations.  For example, if one assigns
prior probabilities $P(H_i)$ to an exhaustive set of hypotheses $H_i$
(say each giving a quantum state of the universe and a set of
operators $E(p)$ in Sensible Quantum Mechanics), and then if the
conditional probability of a perception $p$, given the hypothesis
$H_i$, were $P(p|H_i)$, by Bayes' rule the posterior conditional
probability of the speculation $H_i$ would be
	\begin{equation}
	P(H_i|p)=\frac{P(p|H_i)P(H_i)}{\sum_{j}^{}{P(p|H_j)P(H_j)}}.
	\label{eq:7}
	\end{equation}
Unfortunately, in Sensible Quantum Mechanics one has at most measure
densities $m(p)$ for individual perceptions, and so one cannot
unambiguously give the probability $P(p|H_i)$ without some
specification of the normalization of $m(p)$ in the hypothesis $H_i$.
Possibly one can replace the probability for a perception with the
`typicality' of the perception \cite{Winn,SQM}.  Without some such
solution to this problem, I do not see how to use observations to
turn prior probabilities (say given by some function of the
simplicity of the quantum state and perception operators) into
posterior probabilities for quantum states and perception operators.
Otherwise it would appear that one could only start with a given
quantum state and set of perception operators, and then can one
calculate the measure for all perceptions by Eq.~(\ref{eq:1}) and use
Eq.~(\ref{eq:2}) to calculate conditional probabilities for sets of
perceptions.

\section{Conclusions}

\hspace{.25in}In conclusion, I am proposing that Sensible Quantum
Mechanics, with its Measure Axiom for Perceptions and its Sensible
Quantum Axiom above, is the best framework we have at the present
level for understanding conscious perceptions and the interpretation
of quantum mechanics.  Of course, the framework would only become a
complete theory once one had the set of all perceptions $p$, the
basic measure $dp$ on it, the perception operators $E(p)$, and the
quantum state of the universe.

	Even such a complete theory of perceptions and the physical
world need not be the ultimate simplest complete theory.  There might
be a simpler set of unifying principles from which one could in
principle deduce the perceptions, basic measure, perception
operators, and quantum state, or perhaps some simpler entities that
replaced them.  For example, although in the present framework of
Sensible Quantum Mechanics, the physical world (i.e., the quantum
state), along with the perception operators, determines the measure
density for perceptions in the mental world, there might be a reverse
effect of the mental world affecting the physical world to give a
simpler explanation than we have at present of the correlation
between will and action (why my desire to do something I feel am
capable of doing is correlated with my perception of actually doing
it, i.e., why I ``do as I please'').  If the quantum state is
partially determined by an action functional, can desires in the
mental world affect that functional (say in a coordinate-invariant
way that therefore does not violate energy-momentum conservation)?
Such considerations may call for a more unified framework than
Sensible Quantum Mechanics, which one might call Sensational Quantum
Mechanics.  Such a more unified framework need not violate the
limited assumptions of Sensible Quantum Mechanics, though it might do
that as well and perhaps reduce to Sensible Quantum Mechanics only in
a certain approximate sense.

	To explain these frameworks in terms of an analogy, consider
a classical model of spinless massive point charged particles and an
electromagnetic field in Minkowski spacetime.  Let the charged
particles be analogous to the physical world (the quantum state of
the universe), and the electromagnetic field be analogous to the
mental world (the set of perceptions with its measure $m(p)dp$).  At
the level of a simplistic materialist mind-body philosophy, one might
merely say that the electromagnetic field is part of, or perhaps a
property of, the material particles.  At the level of Sensible
Quantum Mechanics, the charged particle worldlines are the analogue
of the quantum state, the retarded electromagnetic field propagator
(Coulomb's law in the nonrelativistic approximation) is the analogue
of the perception operators, and the electromagnetic field determined
by the worldlines of the charged particles and by the retarded
propagator is the mental world.  (Here you can see that this analogue
of Sensible Quantum Mechanics is valid only if there are no free
incoming electromagnetic waves.)  At the level of Sensational Quantum
Mechanics, at which the mental world may affect the physical world,
the charged particle worldlines are partially determined by the
electromagnetic field through the change in the action it causes.
(This more unified framework better explains the previous level but
does not violate its description, which simply had the particle
worldlines given.)  At a yet higher level, there is the possibility
of incoming free electromagnetic waves, which would violate the
previous frameworks that assumed the electromagnetic field was
uniquely determined by the charged particle worldlines.  Finally, at
a still higher level, there might be an even more unifying framework
in which both charged particles and the electromagnetic field are
seen as modes of a single entity (e.g., to take a popular current
speculation, a superstring).

	Therefore, although it is doubtful that Sensible Quantum
Mechanics is the correct framework for the final unifying theory (if
one does indeed exist), it seems to me to be a move in that direction
that is consistent with what we presently know about consciousness
and the physical world.

\section*{Acknowledgments}

\hspace{.25in}I consciously perceive a gratitude for having had
fruitful discussions or correspondence on this general subject with
David Albert, Andrei Barvinsky, Arvind Borde, David Boulware, Patrick
Brady, Howard Brandt, Bruce Campbell, Brandon Carter, David Deutsch,
Valeri Frolov, Murray Gell-Mann, Farhad Ghoddoussi, Shelly Goldstein,
Richard Gott, Jim Hartle, Geoff Hayward, Bei-Lok Hu, Viqar Husain,
Rocky Kolb, Tom\'{a}\v{s} Kopf, Pavel Krtou\v{s}, Werner Israel, John
Leslie, Andrei Linde, Michael Lockwood, Robb Mann, John Polkinghorne,
David Salopek, Abner Shimony, Euan Squires, Lenny Susskind, Bill
Unruh, Alex Vilenkin, John Wheeler, and various others whose names
aren't in my immediate perception.  I am expecially grateful to my
wife Cathy for helping me become more aware of my feelings.  Finally,
financial support has been provided by the Natural Sciences and
Engineering Council of Canada.


\end{document}